\documentclass[pra,showpacs,onecolumn,superscriptaddress]{revtex4-2}
\usepackage{setup} 

\begin{document}
\title{Multi-sequence alignment using the Quantum Approximate Optimization Algorithm}

\author{Sebastian Yde Madsen}
\affiliation{Department of Physics \& Astronomy, Aarhus University, 8000 Aarhus C, Denmark}
\affiliation{Kvantify Aps, DK-2300 Copenhagen S, Denmark}
\author{Frederik Kofoed Marqversen}
\affiliation{Department of Physics \& Astronomy, Aarhus University, 8000 Aarhus C, Denmark}
\affiliation{Kvantify Aps, DK-2300 Copenhagen S, Denmark}
\author{Stig Elkjær Rasmussen}
\affiliation{Department of Physics \& Astronomy, Aarhus University, 8000 Aarhus C, Denmark}
\affiliation{Kvantify Aps, DK-2300 Copenhagen S, Denmark}
\author{Nikolaj Thomas Zinner}
\affiliation{Department of Physics \& Astronomy, Aarhus University, 8000 Aarhus C, Denmark}
\affiliation{Kvantify Aps, DK-2300 Copenhagen S, Denmark}

\begin{abstract}
The task of \textit{Multiple Sequence Alignment} (MSA) is a constrained combinatorial optimization problem that is generally considered a complex computational problem. In this paper, we first present a binary encoding of MSA and devise a corresponding soft-constrained cost-function that enables a Hamiltonian formulation and implementation of the MSA problem with the variational \textit{Quantum Approximate Optimization Algorithm} (QAOA). Through theoretical analysis, a bound on the ratio of the number of feasible states to the size of the Hilbert space is determined. Furthermore, we consider a small instance of our QAOA-MSA algorithm in both a quantum simulator and its performance on an actual quantum computer. While the ideal solution to the instance of MSA investigated is shown to be the most probable state sampled for a shallow $p<5$ quantum circuit in the simulation, the level of noise in current devices is still a formidable challenge for the kind of MSA-QAOA algorithm developed here. In turn, we are not able to distinguish the feasible solutions from other states in the quantum hardware output data at this point. This indicates a need for further investigation into both the strategy utilized for compiling the quantum circuit, but also the possibility of devising a more compact ansatz, as one might achieve through constraint-preserving mixers for QAOA.
 
\end{abstract}

\maketitle
\section{Introduction}

\begin{spacing}{\myspacing}
A vast variety of present-day technical advances rely heavily on computational models, and their ability to efficiently produce accurate solutions to complex problems. As such, any noticeable improvement in terms of required computational resources, e.g., memory or execution time, for a given problem, can readily be regarded as a very far-reaching and impactful advance to the related scientific fields, and ultimately, everyday technologies.  

Inspired by Paul Benioff's early description of a quantum mechanical Turing Machine (TM) \cite{Benioff1980} and Feynman's 1982 claim that a range of quantum systems can not adequately be simulated on a classical computer \cite{Feynman1982}, David Deutsche, in 1985, claimed that a Universal Turing Machine exists, and that it should be based on a quantum mechanical generalization of the TM, a Quantum TM \cite{Deutsch1985}. This insight was arguably the precursor for the modern field of Quantum Computing, which is in large part concerned with the ability to encode and compute problems, that are classically computationally intractable, through the utilization of quantum mechanical phenomena \cite{Nielsen_Chuang_2022,gill2022quantum,preskill2012quantum}. 

In the field of bio-informatics, one of the most central and computationally challenging problems is \textit{Multiple Sequence Alignment} (MSA), in which $N$ sequences of protein, DNA, RNA etc. are aligned to one another with regard to some predefined scoring metric, in an attempt to gain valuable insights into, i.a., the functional relationship between said sequences \cite{Gagniuc_2021,Mount_2006, Xiong_2006}. 

In the present paper, our aim is to try to estimate how a quantum computer may be used to tackle the MSA problem in bio-informatics. We consider both the abstract model of quantum computing, and the question of encoding of MSA into a set of qubits, as well as the more concrete near-term prospects of implementing such a complex calculation on current hardware. The latter effort is directed at variational quantum algorithms, and more precisely the QAOA approach. Simulations are preformed on classical computers to mimic the behavior of a quantum computer in the absence of noise, and these simulations are used to substantiate that QAOA is in principle capable of handling the MSA problem. However, the noise in present quantum hardware could potentially be detrimental to this prospect. By performing a set of very simple tests on small noisy quantum devices, we confirm this expectation. We conclude our presentation with some remarks on the expected hardware improvements in terms of noise-reduction and circuit depths that would be required to start seeing performance on the MSA problem on par with classical computing approaches. A key observation is that improving the QAOA algorithm to restrict it to a subspace of the full Hilbert space would be required in order to improve algorithmic performance, and that this would again imply the need for larger circuit depths at reducted noise levels.

\end{spacing}

\section{Multiple Sequence Alignment}
\subsection{Biological outline}
\begin{figure}[H]
\tikzset{main node/.style={circle,fill=black!100,draw,minimum size=0.01cm,inner sep=0pt},}
\begin{center}
  \begin{tikzpicture}[decoration={markings,
            mark=between positions 0 and 1 step 0.5
            with { \node[font=\scriptsize,yshift=\pgflinewidth,
                         transform shape] {\ScissorRightBrokenBottom};}}]
    
    \node[] (1) at (2.15,0.4) {A A C G A T T G A};
    \node[] (2) at (1.575,-0.4) {A C A T G A};
    \node[] (3) at (2.8775,-0.475) {\underline{\hspace{0.2cm}}};
    \node[] (4) at (3.2675,-0.475) {\underline{\hspace{0.2cm}}};
    \node[] (5) at (3.6575,-0.475) {\underline{\hspace{0.2cm}}};

        \begin{scope}[blend mode=overlay,overlay]
    \node [label={\sc Unaligned sequences}, style={
      rectangle,
      draw=black,
      thick,
      dashed,
      fill=AU_blue!20,
      opacity=1,
      align=center,
      rounded corners,
      minimum height=2em}, fit= (1) (2) (3) (4) (5), inner sep=0.45cm] {};
    \end{scope}
    \begin{scope}[xshift=6.4cm]
    
    \node[] (1)  at (2.15,0.4)     {A A C G A T T G A};
    \node[] (2)  at (0.625,-0.4)   {A};
    \node[] (3)  at (1.0,-0.475)   {\underline{\hspace{0.2cm}}};
    \node[] (4)  at (1.375,-0.4)   {C};
    \node[] (5)  at (1.75,-0.475)  {\underline{\hspace{0.2cm}}};
    \node[] (6)  at (2.125,-0.4)   {A};
    \node[] (7)  at (2.5,-0.4)     {T};
    \node[] (8)  at (2.875,-0.475) {\underline{\hspace{0.2cm}}};
    \node[] (9)  at (3.25,-0.4)    {G};
    \node[] (10) at (3.625,-0.4)   {A};

    \draw[draw=black,-,line width=0.25mm] (0.625,0.2) -- (0.625,-0.2);
    \draw[draw=black,-,line width=0.25mm] (1.375,0.2) -- (1.375,-0.2);
    \draw[draw=black,-,line width=0.25mm] (2.125,0.2) -- (2.125,-0.2);
    \draw[draw=black,-,line width=0.25mm] (2.5,0.2) -- (2.5,-0.2);
    \draw[draw=black,-,line width=0.25mm] (3.25,0.2) -- (3.25,-0.2);
    \draw[draw=black,-,line width=0.25mm] (3.625,0.2) -- (3.625,-0.2);
    
    \end{scope}
    \begin{scope}[blend mode=overlay,overlay]
    \node [label={\sc Aligned sequences},style={
      rectangle,
      draw=black,
      thick,
      dashed,
      fill=AU_blue!20,
      opacity=1,
      align=center,
      rounded corners,
      minimum height=2em}, fit= (1) (2) (3) (4) (5) (6) (7) (8) (9) (10) , inner sep=0.45cm] {};
    \end{scope}

    \draw[draw=black,-{Straight Barb[width=1.45mm, length=1.5mm]},line width=0.25mm] (4.85,0.0) -- (5.85,0.0);
\end{tikzpicture}    
\end{center}
\vspace{0.3cm}
\caption{One possible Pairwise Alignment of 2 sequences.}
\label{fig: PSA example}
\end{figure}
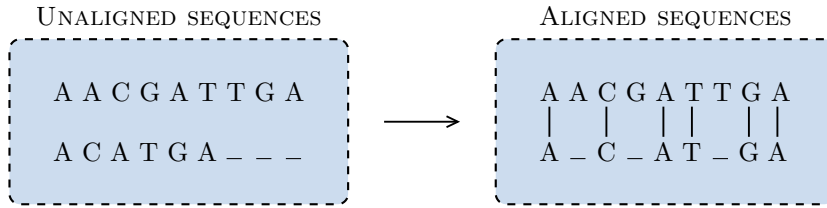

\begin{spacing}{\myspacing}
From a biological point of view, the DNA molecule is of particular interest as it is the structure ultimately responsible for determining a multitude of functional and observable traits of any organism, i.e., the \textit{Phenotype}. DNA sequencing is the process of determining the nucleotide order of a given DNA fragment, and the ability to sequence effectively is of central importance in biological research. Following the implementation of \textit{Next-Generation Sequencing} technologies \cite{NGS-technologies}, the size of available sequence databases has increased many orders of magnitude \cite{SequenceDatabaseSizes}. Therefore, much of the current research done in fields such as Bioinformatics, is concerned with the development and refinement of principal methods such as \textit{sequence alignment}, which are necessary to perform desirable analyses, such as phylogenetic inference, functional analysis
etc. \cite{morrison2006multiple,edgar2006multiple}\\
\textit{Pairwise Sequence alignment} (PSA) of DNA, forms the basis of \textit{Multiple Sequence Alignment} (MSA) and can fundamentally be described as the process of mimicking the naturally occurring mutation process of DNA, by inserting gaps in each of the sequences with the intention of improving the vertical agreement of each column \cite{carroll2006effects,Xiong_2006} (see \cref{fig: PSA example}).
\vspace{1 cm}
\textit{Multiple Sequence Alignment} refers to the process of performing an alignment of three or more sequences.
Formally, the process of sequence alignment is subdivided into the two categories of \textit{local alignment} (LA) and \textit{global alignment} (GA), respectively. As suggested by the name, GA is a form of alignment in which the entirety of the sequences are taken into consideration, whereas LA can be described as the process of determining local areas of interest between the sequences and performing alignment based on these. In the following, we will be concerned with the modeling of GA. 

\subsection{General problem formulation}
Consider a set of $N\geq2$ strings $S = \{s_i\}_{i=1}^N$, each defined on the alphabet $\Lambda=\{A,C,T,G\}$. Let $l_i = |s_i|$ denote the length of each string, and $L\equiv \max\{l_i\}_{i=1}^N$ the length of the longest string. For $1 \leq j \leq l_i$, let $s_{i,j} \in \Lambda$ denote the $j$th letter of the $i$th string, $s_i$. This setup can be illustrated as follows
\begin{figure}[H]
    \begin{center}
        \begin{tikzpicture}
            \node[] (s1) at (0.0, 0.0) {$\vec{S}=
            \begin{cases}
                \vec{s}_1\,\,=(s_{1,1}\,\,\,\,s_{1,2}\,\,\,\,\hdots\,\,\,\,s_{1,l_1})\\
                \vec{s}_2\,\,=(s_{2,1}\,\,\,\,s_{2,2}\,\,\,\,\hdots\,\,\,\,s_{2,l_2})\\
             \quad \quad \quad \quad \quad \quad \quad \quad \vdots \,               \\
                \vec{s}_N  =(s_{N,1} \,\,  s_{N,2} \,\,\, \hdots \,\,\, s_{1,l_N})
            \end{cases}$};

            \node[] (e.g.) at (3.2, 0.0) {e.g.};

            \node[] (s2) at (6.5, 0.0) {$\vec{S}=
            \begin{cases}
                \vec{s}_1\,\,=(C,\,\,C,\,\,T,\,\,G)\\
                \vec{s}_2\,\,=(C,\,\,G)\\
             \quad \quad \quad \quad \vdots \,               \\
                \vec{s}_N  =(C,\,\,T,\,\,G)
            \end{cases}$};
           
            \begin{scope}[blend mode=overlay,overlay]
              \node [style={
              rectangle,
              draw=black,
              thick,
              dashed,
              fill=AU_blue!20,
              opacity=1,
              align=center,
              rounded corners,
              minimum height=2em}, 
              fit= (s2), inner sep=0.15cm] {};
            \end{scope}
            
        \end{tikzpicture}
    \end{center}
\end{figure}

The problem of \textit{Multiple Sequence Alignment} (MSA) is most commonly defined as that of constructing $N$ new strings of equal lengths $l_i=L$ for all $i$, by inserting into each string an appropriate number of gap-characters '$\urule$'.
\begin{equation}
    \begin{tikzpicture}[baseline=0]
        \node[] (s1) at (0.0, 0.0) {$\vec{S}=
        \begin{pmatrix}
            s_{1,1}'\,\,\,\,s_{1,2}'\,\,\,\,\hdots\,\,\,\,s_{1,L}'\\
            s_{2,1}'\,\,\,\,s_{2,2}'\,\,\,\,\hdots\,\,\,\,s_{2,L}'\\
         \quad \quad \, \vdots \,               \\
            s_{N,1}' \,\,  s_{N,2}' \,\,\, \hdots \,\,\, s_{1,L}'
        \end{pmatrix}$};

        \node[] (e.g.) at (3.2, 0.0) {e.g.};

        \node[] (s2) at (6.5, 0.0) {$\vec{S}=
        \begin{pmatrix}
            C&C&T&G\\
            \urule&C&\urule&G\\
             \quad&&\vdots \,               \\
            C&\urule&T&G
        \end{pmatrix}$};
       
        \begin{scope}[blend mode=overlay,overlay]
          \node [style={
          rectangle,
          draw=black,
          thick,
          dashed,
          fill=AU_blue!20,
          opacity=1,
          align=center,
          rounded corners,
          minimum height=2em}, 
          fit= (s2), inner sep=0.15cm] {};
        \end{scope}
        
    \end{tikzpicture}
    \label{eq: standard MSA matrix}
\end{equation} 
The only immediate restriction on a given MSA solution is therefore that it has to preserve the order of the characters. The version of MSA described above does not in general exclude the possibility that two of the initial strings both have maximal length. However, we restrict ourselves to the slightly less general case where exactly one string has maximal length. This string can be viewed as the reference string since it is fixed to its initial ordering.

Note that the exact formulation of a specific instance of MSA is highly dependent upon the biological nature of the problem. If the purpose of the MSA is to perform evolutionary reconstruction, e.g., create a \textit{Phylogenetic tree} of the sequences, then the accuracy of the MSA is highly dependent on the aligned areas of the sequences being homologous. On the other hand, if the MSA is supposed to demonstrate structural equality, the aligned areas of the sequences ought to have comparable positioning in their respective 2 or 3D structures \cite{Chatzou2016}. 

\subsection{Scoring strategies}
Currently, a vast multitude of different scoring schemes exist, both in terms of the computation of intermediate scores in the recursively defined matrices in the \textit{Dynamical Programming}
(DP) approach, and in terms of scoring some given MSA.  Each of these schemes emphasizes different biological features and as such favor different alignments. The \textit{Sum-of-Pairs} (SP) scoring scheme is one of the most common schemes classically implemented, which assigns a scoring based on the degree of vertical similarity between the sequences, post alignment.
As such, given an MSA of the form shown in \cref{eq: standard MSA matrix}, the SP scheme is defined as the sum of all distinct pairwise comparisons in a column, over all columns
\begin{equation}
    \textrm{SP}(\vec{S}')=\sum_{i=1}^{L}\sum_{j=1}^{N}\sum_{j'=j+1}^{N}\textrm{sim}_{\textrm{SP}}(s_{j,i}\,,\,s_{j',i}),
\end{equation}
with a native similarity score defined as:
\begin{equation}
    \textrm{sim}_{\textrm{SP}}(s_{j,i}\,,\,s_{j',i}) = 
    \begin{cases}
    -1, \quad s_{j,i} = s_{j',i} \\
    +1, \quad s_{j,i} \neq  s_{j',i} \\
    \phantom{+}0, \quad s_{j,i} =\textrm{'}\urule\textrm{'}\,\, \lor\,\, s_{j',i} =\textrm{'}\urule\textrm{'}
    \end{cases} .
\end{equation}
\subsection{Intractability of MSA \& current approaches}
By a combinatorial analysis (see \cref{sec: Combinatorial analysis of MSA}) it becomes apparent that MSA too suffers from the 'curse of dimensionality', meaning that the number of problem configurations intrinsically expands in at least an exponential manner. Even for the restricted subclass of MSA with an initially fixed number of gaps available for insertion in each sequence, the search space $\mathcal{S}$ is intractably large. In particular, aligning a mere 10 query sequences against a target sequence of 50 characters with 7 gaps available for insertion in each, the number of possible alignments becomes comparable with the number of atoms in the known universe:
\begin{equation}
    \big|\mathcal{S}\big| = \binom{50}{7}^9\sim10^{79}.
\end{equation}
As the fastest supercomputer is currently able to perform $\sim 10^{18}$ Flop$/$s \cite{top500}, utilizing a brute-force type of search algorithm to generate and evaluate all $\sim {10^{79}}$ possible MSA's would still require a completely intractable amount of time. This firmly settles the fact that any such approach is out of the scope of any classical algorithm, assuming of course that one is concerned with a guarantee of finding the global minimum of the chosen scoring metric. This observation is further supported by the fact that MSA is known to be an NP-hard problem under various scoring metrics, including SP \cite{BONIZZONI200163, just2001computational}.

\end{spacing}

\subsubsection{Classical approaches to MSA}

\begin{spacing}{\myspacing}
As PSA forms the basis of MSA, several algorithms have been developed in the past decades \cite{Chao2022,Chatzou2016,Thompson2011}, in the attempt to provide accurate alignments of multiple sequences efficiently, by initially optimizing the process of aligning a pair of sequences. Most notable instances include the classical DP approach of the \textit{Needleman–Wunsch} (NW) algorithm for global alignment \cite{Needleman1970} and \textit{Smith-Waterman} (SM) algorithm for local alignment \cite{Smith1981}. For PSA of strings with lengths $\{L_1,L_2\}$, both NW and SM is of time and space complexity $\mathcal{O}(L_1L_2)$\cite{Jararweh2019}, which, for the restricted MSA with $N$ fixed-length strings generalizes to $\mathcal{O}(L^N)$\cite{Carrillo1988}.
\end{spacing}

\section{Encodings}
\begin{spacing}{\myspacing}
To solve the combinatorial problem of finding the best MSA on a quantum computer, the problem has to be encoded in a way that can be represented by a quantum state. Typically, a binary encoding is chosen such that any given binary string, $z\in\{0,1\}^n$, can be represented by a product state in the \textit{Pauli-Z} basis, e.g.: $001100 \equalhat \ket{001100}$. The task of defining a valid encoding effectively amounts to finding a binary encoding of the problem which is expressive enough \cite{sim2019expressibility,rasmussen2020reducing} that it can represent any state for a given problem size, but on the other hand, utilizes the shortest possible binary string, to avoid redundancy and computational overhead.
\subsection{One-Hot Column encoding}
\begin{figure}[H]
\vspace{1cm}
    \begin{center}
    \scalebox{0.97}{\begin{tikzpicture}

    \draw (-1.5,1.4) -- (0,1.4) -- (0,0) -- (-1.5,0) -- (-1.5,1.4);
    
    \node[] (s1) at (-0.75, 1.1) {$A\,C\,G\,G$};
    \node[] (ss1) at (-1.95, 1.1) {$s_1:$};
    
    \node[] (s2) at (-0.75, 0.7) {$A\,G\,G\,\urule$};
    \node[] (ss2) at (-1.95, 0.7) {$s_2:$};
    
    \node[] (s3) at (-0.75, 0.3) {$A\,T\,\urule\,\urule$};
    \node[] (ss3) at (-1.95, 0.3) {$s_3:$};
    
    \begin{scope}[blend mode=overlay,overlay]
    \node [label={\sc Original sequences}, style={
      rectangle,
      draw=black,
      thick,
      dashed,
      fill=AU_blue!20,
      opacity=1,
      align=center,
      rounded corners,
      minimum height=2em}, fit= (s1) (ss1) (s2) (ss2) (s3) (ss3), inner sep=0.3cm] {};
    \end{scope}
    
    \draw[-stealth] (0.1,0.7) -- (0.9,0.7);
    
    \draw (-1.5+2.5,1.4) -- (2.8+2.5,1.4) -- (2.8+2.5,0) -- (-1.5+2.5,0) -- (-1.5+2.5,1.4);
    
    \node[] (encoding1) at (-0.75+3.9, 1.1) {$\ket{1000}\,\ket{0100}\,\ket{0010}\,\ket{0001}$};
    \node[] (encoding2) at (-1.275+3.9, 0.7) {$\ket{1000}\,\ket{0100}\,\ket{0010}$};
    \node[] (encoding3) at (-1.78+3.9, 0.3) {$\ket{1000}\,\ket{0100}\,$};

    \begin{scope}[blend mode=overlay,overlay]
    \node [label={\sc Encoding}, style={
      rectangle,
      draw=black,
      thick,
      dashed,
      fill=AU_blue!20,
      opacity=1,
      align=center,
      rounded corners,
      minimum height=2em}, fit= (encoding1) (encoding2) (encoding3), inner sep=0.3cm] {};
      
    \end{scope}
    \draw[-stealth] (5.4,0.7) -- (6.1,0.7);
    
    \node[] (finalstring0) at (6.28,0.71){$|$};
    \node[] (finalstring) at (9.8,0.65) {$\underbrace{\strut\overbrace{\strut1000}^{n_1}\overbrace{\strut0100}^{n_2}\overbrace{\strut0010}^{n_3}\overbrace{\strut0001}^{n_4}}_{s_1}\underbrace{\strut\overbrace{\strut1000}^{n_1}\overbrace{\strut0100}^{n_2}\overbrace{\strut0010}^{n_3}}_{s_2}\underbrace{\strut\overbrace{\strut1000}^{n_1}\overbrace{\strut0100}^{n_2}}_{s_3}\rangle$};
    
    \begin{scope}[blend mode=overlay,overlay]
    \node [label={\sc Corresponding state}, style={
      rectangle,
      draw=black,
      thick,
      dashed,
      fill=AU_blue!20,
      opacity=1,
      align=center,
      rounded corners,
      minimum height=2em}, fit= (finalstring0) (finalstring), inner xsep=0.3cm, inner ysep=0.3] {};
    \end{scope}
    
    \end{tikzpicture}}
    \end{center}
    \caption{A generic example showing the \textit{One-Hot column} encoding chosen for this instance of MSA}
    \label{fig: encoding example}
\end{figure}
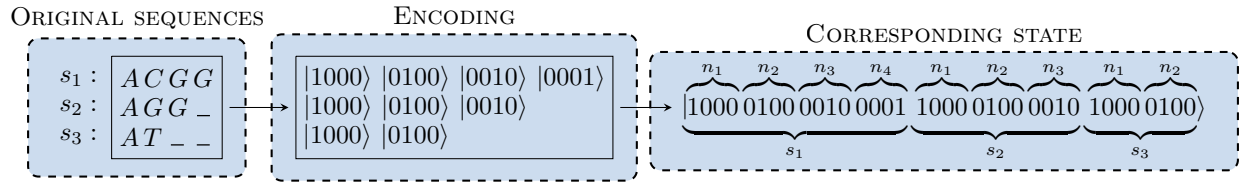

A straightforward way of encoding any $N$ strings of lengths $(l_i)_{i=1}^N$, arranged in a matrix, is by considering if the $n$th character in the $s$th string is placed in the $i$th column of the matrix, such that each letter would require a binary string of length corresponding the width of the matrix. Specifically, the character $A$ in the string $ACGG$ would be encoded as $1000$, as it would inhabit the first column of the matrix. A full example of the One-Hot Column encoding is illustrated in \cref{fig: encoding example} from which it can be seen that the final state, encoding the sequence of strings, is given as a concatenation of the individual 'encoding segments'.
As the number of columns in the matrix corresponds to the length of the longest string (the \textit{target string}), this encoding generally requires:
\begin{equation}
   n= L\sum_{i=1}^Nl_i,\quad \textrm{with}\quad L\equiv \max\{l_i\}_{i=1}^N,
\end{equation}
qubits to encode, such that the number of qubits scales polynomially in the maximum string length $n=\mathcal{O}(NL^2)$.

\subsection{Cost function}
\subsubsection{Hard constraints}
Naturally, almost any encoding enforces a handful of constraints specific to the problem at hand. The One-Hot Column encoding chosen for MSA, entails the following \textit{hard constraints} on feasible solutions: 
\begin{itemize}
    \item Any $n$th letter in any $s$th string must occur at exactly one $i$th of the $L$ available columns:
    \begin{equation}
        \sum_{i=1}^{L}x_{s,n,i}=1.
            ,\quad
        \forall s,n
    \end{equation}
    \item Any $i$th of the $L$ available columns, in each row, must be occupied by one letter at most:
    \begin{equation}
        \sum_{n=1}^{l_s}x_{s,n,i}\leq1.
            ,\quad
        \forall s, i
    \end{equation}
    where $l_s$ denotes the length of the $s$th string.
\end{itemize}
Furthermore, any MSA has to preserve the initial order of letters in each of the strings, and as such, can be seen as the most, according to some arbitrary scoring metric, beneficial ordering of gaps between the letters. This formulation of MSA inherently imposes an additional \textit{hard constraint}
\begin{itemize}
    \item No $(n+1)$th letter in a string must be placed in a column before the $n$th letter:
    \begin{equation}
        \sum_{s=1}^{N}x_{s,n',i} x_{s,n,i'} = 0
            ,\quad
        \forall n<n',\,\forall i<i'.
    \end{equation}

\end{itemize}

\subsubsection{Scoring scheme}
As the described $x_{s,n,i}$ encoding, alongside the introduced SP-scoring scheme, is incapable of distinguishing different letters of the defined alphabet, $\Lambda$, the scheme cannot directly be implemented in the cost function. A possible native way around this issue is an implementation in which all relevant pair-wise scorings of characters in $\vec{S}$, are encoded as weights in matrices, as a part of the pre-processing routine. Specifically, for each distinct pair of sequences in a matrix, one would calculate a matrix $[\omega_{i,j}]\equiv \textrm{score}(s_{s_1,i}\,,\,s_{s_2,j})$. As such, preparing $\mathbf{\Omega}$, allows for a scoring equivalent to the SP-scheme to be implemented as:
\begin{equation}
    \sum_{s<s'}\sum_{n,n}\sum_{i}\omega_{s,n,s',n'}x_{s,n,i}x_{s',n',i},
\end{equation}
in which $\omega_{s,n,s',n'}$ then denotes the predefined scoring associated with alignment of the $n$th character in the $s$th string with the $(n')$th character of the $(s')$th string.

\subsubsection{Soft constraints}
For the resulting cost function to be applicable, the aforementioned \textit{hard constraints} has to be transformed to analogous \textit{soft constraints}, associating each term with some predefined factor, penalizing states in violation of the constraints. In general, this is not a trivial task but Ref. \cite{QUBOtutorial} provides a few constraint-penalty pairs, from which the following was inferred:
\begin{align}
    \sum_{i}x_{s,n,i} &= 1, \quad \forall s,n & &\Longrightarrow & &p_1\sum_s\sum_n\bigg(\sum_ix_{s,n,i}-1\bigg)^2,
        \\ 
    \sum_{n}x_{s,n,i} &\leq 1, \quad \forall s,i & &\Longrightarrow & &p_2\sum_s\sum_i\sum_{n<n'}x_{s,n,i}x_{s,n',i},
        \\
    \sum_{s}x_{s,n',i}x_{s,n,i'} &= 0, \quad \forall n<n',\,\forall i<i' & &\Longrightarrow & &p_3\sum_s\sum_{n<n'}\sum_{i<i'}x_{s,n',i}x_{s,n,i'} ,
\end{align}
yielding a final expression for the cost of any binary encoding of some MSA matrix $x\in\{0,1\}^n$ as:
\begin{equation}
\begin{aligned}
    \mathcal{C}(x) &=     \sum_{s<s'}\sum_{n,n'}\sum_{i}\omega_{s,n,s',n'}x_{s,n,i}x_{s',n',i} +p_1\sum_s\sum_n\bigg(\sum_ix_{s,n,i}-1\bigg)^2\\
    &+p_2\sum_s\sum_i\sum_{n<n'}x_{s,n,i}x_{s,n',i}
    +p_3\sum_s\sum_{n<n'}\sum_{i<i'}x_{s,n',i}x_{s,n,i'}.
    \label{eq: cost function}
\end{aligned}
\end{equation}

\subsection{Hamiltonian}
\noindent The fundamental objective is to encode the cost in the Hamiltonian operator, diagonal in the computational \textit{Pauli-Z} basis $\mathcal{C}(x)\longrightarrow\hat{H}(|x\rangle)$ such that the eigenvalues of different feasible MSAs corresponds to eigenvalues of the Hamiltonian:
\begin{equation}
    \hat{\mathcal{H}}\ket{x} = \mathcal{C}(x)\ket{x},
    \label{eq:hamiltonian eigenequation}
\end{equation}
and as the \textit{Pauli-Z} operator has eigenvalues $\{-1,+1\}$
\begin{equation}
    \hat{\sigma_z}\,|y\rangle = (-1)^{y}|y\rangle,\quad y\in\{0,1\}.
\end{equation}
The process of encoding the cost in $\hat{H}$, amounts to the transformation:
\begin{equation}
    x_{s,n,i}\rightarrow\frac{\mathbbm{1}-\hat{\sigma_z}^{s,n,i}}{2},
    \label{eq: bin to z}
\end{equation}
in which $\hat{\sigma_z}^{s,n,i}$ corresponds to the \textit{Pauli-Z} operator acting on the $(s,n,i)$th qubit tensor-product state corresponding to the binary encoding of the given MSA matrix. In practice, this is accomplished by initially casting the cost function into a \textit{Quadratic Unconstrained Binary Optimization} (QUBO) model, of the form:
\begin{equation}
    C(\vec{x})=\vec{x}^T\vec{Q}\vec{x}+\vec{x}^T\vec{h}+d,\quad \vec{x}\in\{0,1\}^n,
    \label{eq: general QUBO}
\end{equation}
and then mapping the QUBO model to an Ising model.
Consequently, this is achieved by decomposing the vectorized QUBO model into a sum and imposing the variable change resulting from \cref{eq: bin to z}, which yields:
\begin{equation*}
    C(\vec{s})=\underbrace{\frac{1}{4}\vec{s}^T\vec{Q}\vec{s}}_{\text{Quadratic term}}-\,\,\,\underbrace{\frac{1}{4}\bigg({\mathbbm{1}_n}^T\vec{Q}^T+{\mathbbm{1}_n}^T\vec{Q}+2\vec{h}^T\bigg)\vec{s}}_{\text{Linear term}}\,\,\,+\,\,\,\underbrace{\frac{1}{4}{\mathbbm{1}_n}^T\vec{Q}{\mathbbm{1}_n}+\frac{1}{2}{\mathbbm{1}_n}^T\vec{h}+d}_{\text{Constant term}},
\end{equation*}
in which ${\mathbbm{1}_n}=\{1\}^n$ is a $n$-length vector of ones (See \cref{sec: QUBO to ising} for a more explanatory review of the mathematical transformation). \\
Specifically, the $(i,j)$th off-diagonal entries in the quadratic term represent the coupling strength between qubit $i$ and $j$, and is used as an estimator for the rotation angle in the unitary exponentiation of the two-qubit \textit{Pauli-Z} term in the Hamiltonian acting on qubit $i$ and $j$, whilst the $i$th diagonal entry, together with the $i$th entry in the linear term is used as an estimator of the angle used in the unitary exponentiation of the single-qubit \textit{Pauli-Z} term acting on the $i$th qubit.

\end{spacing}

\section{Variational Quantum Algorithms}
\begin{spacing}{\myspacing}

Variational Quantum Algorithms (VQAs) are one of the leading strategies within the noisy intermediate-scale quantum (NISQ) era \cite{preskill2018quantum,lau2022nisq}. VQAs seek to provide a very general body for efficiently dealing with computationally complex tasks, which is achieved by means of hybrid quantum-classical (HQC) algorithms, where a classical computer is used to variationally update the variables $\boldsymbol{\theta}$ of a parameterized quantum circuit (PQC), thus attempting to improve upon a purely classical computation, by delegating specific sub-processes to a quantum computer \cite{Cerezo_2021}. The underlying principle of VQAs is somewhat inspired by the 'classical' \textit{Variational Method} of quantum mechanics that approximates the ground state energy, $E_0$, of some system by initially defining some trial state (oftentimes termed \textit{ansatz}) as a function of some parameters, and then minimizes the expectation of the Hamiltonian with respect to these parameters, with the intend of approving upon the approximation of $E_0$ \cite{blekos2023review}. 

\subsection{Working principle}
In the type of VQA considered here, the problem-specific cost function $\mathcal{C}(\boldsymbol{\theta})$ is mapped to an equivalent Hamiltonian $\hat{H}$ which is then encoded into the PQC, such that the action of the cost function on any viable set of parameters, corresponds to 'passing' some initial state $|\psi_{0}\rangle$ through the PQC. Prior to measurement, parts of the circuit might be transformed to the appropriate measurement basis and the estimate of $\langle\hat{H}\rangle(\boldsymbol{\theta})$ is determined by repeated sampling of the circuit, according to suitable statistical requirements of certainty. Finally, the classical optimizer navigates the cost landscape encoded by the Hamiltonian in the PQC, to determine the parameters $\boldsymbol{\theta^{\ast}}$ corresponding to the minimum of the problem-specific cost function:
\begin{equation}
    \boldsymbol{\theta^{\ast}}=\argmin_{\boldsymbol{\theta}}\big( \mathcal{C}(\boldsymbol{\theta})\big).
\end{equation}
The final state of the system is determined upon successive repetitions of sampling the PQC and performing classical minimization, until some predetermined criteria of accuracy are met. Now, as the PQC ultimately corresponds to some unitary transformation $\hat{U}(\boldsymbol{\theta})$, this final state is given as:
\begin{equation}
    |\psi(\boldsymbol{\theta})\rangle = \hat{U}(\boldsymbol{\theta})|\psi_{0}\rangle.
\end{equation}

\subsection{Quantum Approximate Optimization Algorithm}
\begin{figure}[H]
    \centering
    \includegraphics[width=\linewidth]{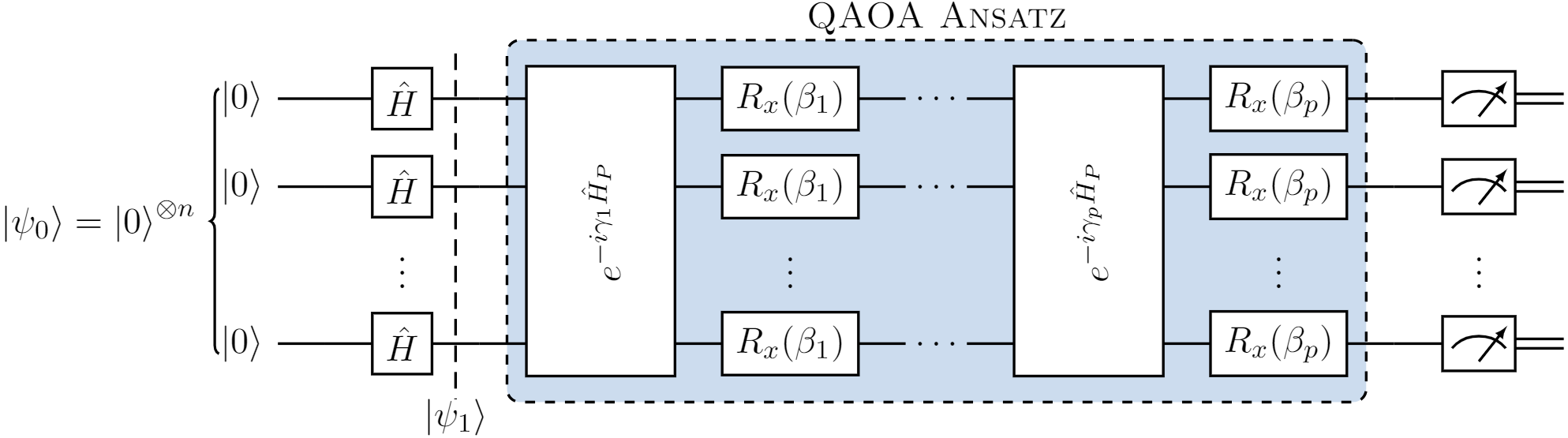}
    \caption{A generic depiction of the repeating layers of cost and mixer hamiltonians in
the QAOA ansatz.}
    \label{fig: QAOA ansatz}
\end{figure}

The Quantum Approximate Optimization Algorithm (QAOA), initially proposed by ref. \cite{farhi2014qaoa}, is a VQA designed for determining approximate solutions to combinatorial optimization problems, inspired by a trotterized version of the Quantum Annealing Algorithm \cite{farhi2000quantum,farhi2001quantum}.
Originally, the QAOA ansatz was proposed as a repetitive alternation between a problem-specific parameterized unitary and a parameterized 'mixer' unitary. The problem-specific unitary $\hat{U}_P(\gamma)=\e^{-\i\gamma\hat{H}_P}$ is defined by a Hamiltonian
\begin{equation}
    \hat{H}_P = \sum_{j}^Kc_j\hat{P}_j^z,\quad \hat{P}_j^z\in\{\hat{\sigma}_z,\hat{\sigma}_z\otimes\hat{\sigma}_z,...,\hat{\sigma}_z^{\otimes n}\},
\end{equation}
diagonal in the space of feasible solutions for the given problem. The cost corresponding to a solution state is encoded in the eigenvalue $\hat{H}_P|\boldsymbol{z}\rangle=\mathcal{C}(\boldsymbol{z})|\boldsymbol{z}\rangle$ of that state. The parameterized 'mixer' unitary $\hat{U}_M(\beta)=\e^{-\i\beta\hat{H}_M}$, is generated by single qubit bit-flips
\begin{equation}
\e^{-\i\beta\hat{H}_M}=\big(R_x(\beta/2)\big)^{\otimes n}.
\end{equation}
Preparing the input in the uniform superposition
\begin{equation}
    |\psi_1\rangle=\hat{H}^{\otimes n}|0\rangle^{\otimes n}=\bigg(\frac{1}{\sqrt{2}}|0\rangle+|1\rangle\bigg)^{\otimes n}=\frac{1}{\sqrt{2^n}}\sum_{x\in\{0,1\}^n}|x\rangle,
\end{equation}
then produces the following trial state:
\begin{equation}
    |\psi(\boldsymbol{\beta},\boldsymbol{\gamma})\rangle=\underbrace{\bigg(\e^{-\i\gamma_1\hat{H}_P}\e^{-\i\beta_1\hat{H}_M}\bigg)\bigg(\e^{-\i\gamma_2\hat{H}_P}\e^{-\i\beta_2\hat{H}_M}\bigg)\hdots\bigg(\e^{-\i\gamma_p\hat{H}_P}\e^{-\i\beta_p\hat{H}_M}\bigg)}_{\textrm{$p$ repetitions}}|\psi_1\rangle.
\end{equation}
In conclusion, these steps come together resulting in the circuit shown in \cref{fig: QAOA ansatz}.

After an appropriate number of shots the VQA strategy would be accompanied by a subroutine of classical optimization, determining: 
\begin{equation}
    \bm{\beta}',\bm{\gamma'} = \argmin_{\bm{\beta},\bm{\gamma}\in\R^p} \big(\langle \psi(\bm{\beta},\bm{\gamma})|\hat{H}_P|\psi(\bm{\beta},\bm{\gamma})\rangle\big),
\end{equation}
and setting:
\begin{equation}
    \hat{U}(\boldsymbol{\beta},\boldsymbol{\gamma})\to\hat{U}(\boldsymbol{\beta}',\boldsymbol{\gamma}'),
\end{equation}
prior to the computation of the next trial state. This effectively means that the circuit depth increases as $\mathcal{O}(p)$ since each repetition of the mixer unitary and Hamiltonian unitary, adds a fixed number of extra gates. From a theoretical point of view, the cardinal advantage of the QAOA stems from the fact that repeated applications of the mixer and the cost unitary yield a high expressivity, such that there exists some $\boldsymbol{\beta}^{\ast},\boldsymbol{\gamma}^{\ast}$ for which:
\begin{equation}
    \lim_{p\to\infty}\big(\langle \psi(\boldsymbol{\beta}^{\ast},\boldsymbol{\gamma}^{\ast})|\hat{H}_P|\psi(\boldsymbol{\beta}^{\ast},\boldsymbol{\gamma}^{\ast})\rangle\big) = \mathcal{C}_\textrm{min},
    \label{eq: theoretical guarantee}
\end{equation}
This means that, provided $\boldsymbol{\beta}^{\ast}$ and $\boldsymbol{\gamma}^{\ast}$ can be efficiently determined, the optimal solution $z^{\ast}\in\{0,1\}^n$ can be determined by means of QAOA and corresponds to the bit string encoded in $|\psi(\boldsymbol{\beta}^{\ast},\boldsymbol{\gamma}^{\ast})\rangle$ \cite{farhi2014qaoa}.

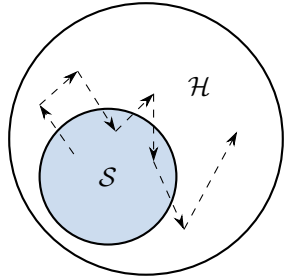
\begin{figure}[H]
\tikzset{cross/.style={cross out, draw=black, minimum size=2*(#1-\pgflinewidth), inner sep=0pt, outer sep=0pt},cross/.default={1.35pt}}
\begin{center}
\begin{tikzpicture}
\draw[color=black, thick](0,0) circle (1.8);
\filldraw[color=black, fill=AU_blue!20,  thick](-0.5,-0.5) circle (0.9);
\node[] at (-0.5,-0.5) {$\mathcal{S}$};
\node[] at (0.7,0.7) {$\mathcal{H}$};
\draw[dashed,-{Stealth[length=2mm, width=1.25mm]}] (-0.95, -0.2) -- (-1.4, 0.45);
\draw[dashed,-{Stealth[length=2mm, width=1.25mm]}]  (-1.4, 0.45)  -- (-0.9, 0.9);
\draw[dashed,-{Stealth[length=2mm, width=1.25mm]}]  (-0.9, 0.9)   -- (-0.4, 0.1);
\draw[dashed,-{Stealth[length=2mm, width=1.25mm]}]  (-0.4, 0.1)   -- (0.1, 0.6);
\draw[dashed,-{Stealth[length=2mm, width=1.25mm]}]  (0.1, 0.6)   -- (0.1, -0.3);
\draw[dashed,-{Stealth[length=2mm, width=1.25mm]}] (0.1, -0.3)   -- (0.5, -1.2);
\draw[dashed,-{Stealth[length=2mm, width=1.25mm]}]  (0.5, -1.2)   -- (1.2, 0.1);
\end{tikzpicture}
\end{center}
\caption{Venn diagram showing the subspace of feasible solutions, $\mathcal{S}$, being smaller than the Hilbert space $\mathcal{H}$ of all $2^n$ states, accompanied by a symbolic depiction of the ineffective search of feasible solutions.}
\label{fig: heh}
\end{figure}

This effectively means that the mixer-terms can be seen as the part of the ansatz assuring that the algorithm is capable of searching the entire Hilbert space, $\mathcal{H}$.
However, as the size of $\mathcal{H}$ generally grows exponentially with respect to the number of qubits, $n$, encoding the problem, i.e., $|\mathcal{H}|=2^n$, whereas the set of feasible solutions $\mathcal{S}$ to a given problem might only constitute a fraction, i.e., $\mathcal{S}\subseteq \mathcal{H}$. The generic \textit{Pauli-X} mixer chosen in the standard QAOA ansatz suffers from the problem of potentially allowing for the evaluation of many irrelevant states, see figure~\ref{fig: heh}.

It can be seen, that for the specific formulation of MSA and encoding chosen in this project, the ratio, for any $N\geq2$ strings of general integer lengths $(l_i)_{i=1}^N,\,\, l_i\in\mathbbm{N}$, with a unique maximum length, such that 
\begin{equation}
    \textrm{if}\quad L \equiv l_j\equiv\max\{l_i\}_{i=1}^N\quad\textrm{then}\quad l_j\neq l_i\Leftrightarrow i\neq j,
\end{equation}
is upper bounded by a function that is super-exponentially decreasing in $L$ and exponentially decreasing in $N$:
\begin{equation}
   \frac{|\mathcal{S}|}{|\mathcal{H}|}\leq \frac{1}{L!} e^{-[\ln(2)L-\ln(L)][L+N-1]},\quad  \forall\, N,L\in \N_{\geq 2}.
\end{equation}
See Appendix \ref{sec: Combinatorial analysis of MSA} for details.

This strongly suggests that the encoding chosen for the problem, has a lot of redundancy and that the given combination of encoding and constraints might benefit from specialized constraint-preserving mixers that limit the search to the feasible subspace \cite{Hadfield_2019, Fuchs_2022}. 

\subsubsection{QAOA Toy Model for MSA}
\label{sec: MSA toy model for QAOA}

Consider the following native 2-sequence example of MSA:
\begin{equation}
    \vec{S}=
    \ptrix{A \, G \\ 
    G \, \urule}=
    \ctrix{\ket{x_{111}x_{112}} & \ket{x_{121}x_{122}} \\ 
    \ket{x_{211}x_{212}} & \urule}
    \equalhat
    \ket{x_{111}x_{112}x_{121}x_{122}x_{211}x_{212}}
    =\ket{100110},
    \label{eq: Simple MSA toy model}
\end{equation}
with the previously mentioned weight matrix for the two rows, filled according to the SP scheme:
\begin{equation}
    \Omega_{s_1,s_2}=\ctrix{\omega_{1121} & \omega_{1122}\\
    \omega_{1221} & \omega_{1222}}=\ctrix{\,\,\,1 & 0 \\ -1 & 0}.
\end{equation}
The implementation of the cost function defined in \cref{eq: cost function}, in the QAOA ansatz for a circuit of layer repetition $p=1$, for the case given in \cref{eq: Simple MSA toy model} results in the circuit seen on \cref{fig:circuit} in \cref{sec: circuit}.\\

\begin{figure}[H]
     \centering
     \includegraphics[width=\linewidth]{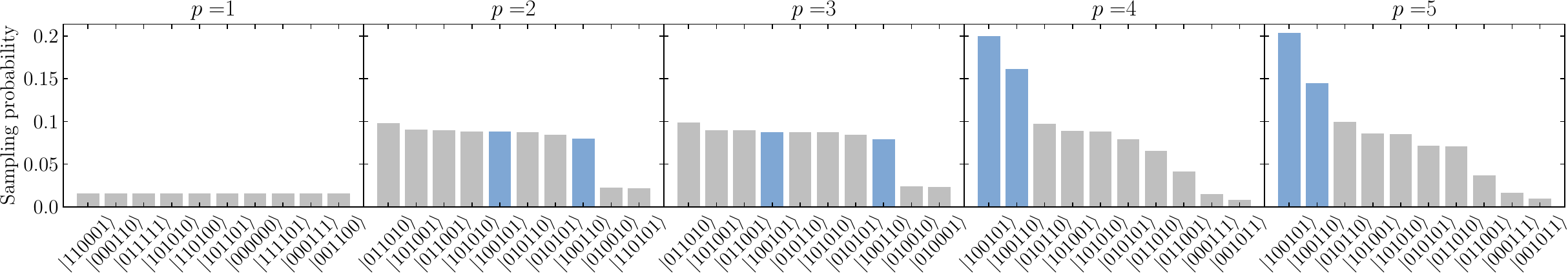}
        \caption{Histograms depicting the 10 highest probabilities resulting from the simulation of \cref{eq: Simple MSA toy model}, for a QAOA circuit consisting of layer repetitions $p=1,2,\hdots,5$, for 5000 shots. The penalties were set to $p_1=10,\, p_2=p_3=1$ in each of the instances. The \textit{blue} bars correspond to feasible solutions.}
        \label{fig: histogramsfirst}
\end{figure}

The probability distributions resulting from simulations of this circuit, for an increasing number of repetitions, utilizing Qiskit's framework, alongside the 'classical' Constrained Optimization by Linear Approximation (COBYLA) algorithm \cite{powell1994direct,powell1998direct,powell2007view}, as implemented by ref. \cite{2020SciPy-NMeth}, chosen for the parameter optimization subroutine is shown on \cref{fig: histogramsfirst}.

\begin{figure}[H]
    \centering
\includegraphics[width=0.27\linewidth]{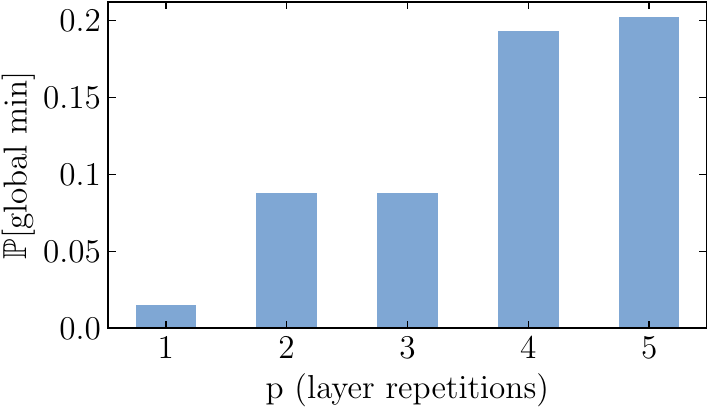}
    \caption{Histogram depicting the probability of sampling the global minimum $\ket{100101}$ as a function of the number of repetitions $p$, resulting from the simulation of \cref{eq: Simple MSA toy model}, for a QAOA circuit, with  5000 shots taken. The penalties were set to $p_1=10,\, p_2=p_3=1$ in each of the instances. }
    \label{fig: global min increase}
\end{figure}

From the figure, it can be seen that the probability of sampling both the ideal alignment (the global minimum, corresponding to $\ket{100101}$), and, in general, valid alignments, improves, as the number of layer repetitions is increased, and specifically, that the probability of sampling a state corresponding to the global minimum, increases with the number of repetitions, as seen on \cref{fig: global min increase}.

\begin{figure}[H]
    \centering
\includegraphics[width=\linewidth]{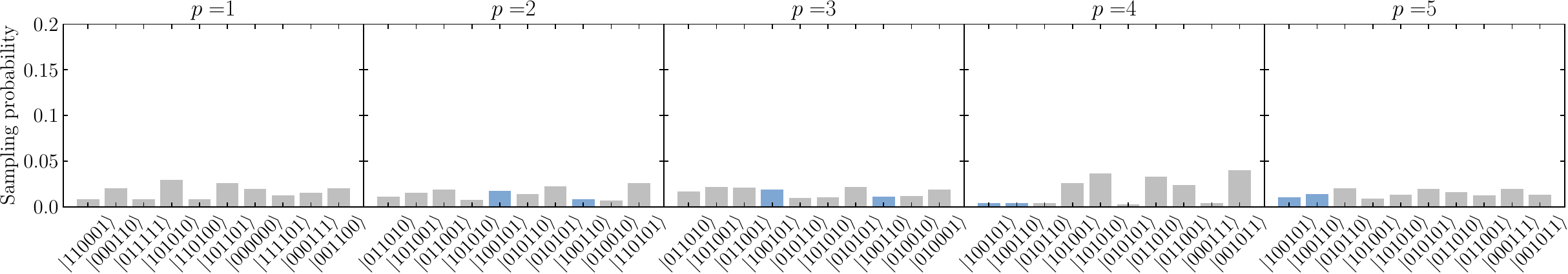}
    \caption{Histograms depicting sampling probabilities, resulting from running \cref{eq: Simple MSA toy model}, for a QAOA circuit consisting of layer repetitions $p=1,\,p=2,\hdots,p=5$, for 5000 shots taken, on the Rigetti Aspen-M-3 superconducting quantum computer. The penalties were set to $p_1=10,\, p_2=p_3=1$ in each of the instances. It should be noted that the states are ordered, not with respect to sampling probability, but in the same manner as in \cref{fig: global min increase} for comparability. For that reason, the most probable states resulting from these runs, are in general not guaranteed to be visible in the figure. The \textit{blue} bars correspond to feasible solutions.}
    \label{fig: global min increase real}
\end{figure}

Having these ideal and noiseless results as our baseline, we now proceed to consider how the algorithm performs on actual quantum hardware. In particular, we ran tests on the 
on Rigetti's Aspen-M-3 79 qubit superconducting quantum computer. The resulting distributions from the quantum hardware are shown in \cref{fig: global min increase real}.

As evident from the seemingly unordered distributions in \cref{fig: global min increase real}, the noise level of currently available hardware might still be too significant to enable high probability sampling of feasible solutions for the MSA QAOA type algorithm designed here. Though, it should be noted that the specific choice of circuit-compilation procedure \cite{maronese2022quantum}, and the chosen subset of qubits for a specific topology influences performance \cite{pelofske2022quantum}, such that better results might very well be achieved through the application of another and more specialized compiler. 

\end{spacing}

\section{Conclusion}

\begin{spacing}{\myspacing}
By considering Multiple Sequence Alignment as a constrained combinatorial optimization problem, we demonstrated that it is indeed possible to design a binary encoding for the investigated instance of MSA, such that there can be found a 'faithful' cost function which, in turn, can be implemented in a near-term ansatz on a gate based quantum computer. 

However, for the proposed near-term QAOA method, a variety of challenges arise. Firstly it is seen that, generally, implementing a cost function with soft constraints, entails additional hyper-parameters in the form of the penalty factors, which in turn has to be optimized, alongside the required number of layer repetitions $p$, as a part of the pre-processing prior to execution and sampling of the circuit. Furthermore, as near-term quantum devices are prone to noise, any meaningful encoding, and corresponding cost function, has to be designed in a manner that entails sparse requirements to the quantum circuit in terms of gates and qubits required for computation, to avoid faulty computation and the resulting measurement overhead. The specific \textit{one-hot column} encoding chosen for MSA displays heavy redundancy as only an exponentially decreasing fraction of the $2^n$ different states, representable by the encoding, corresponds to feasible rearrangements of the letters in the MSA matrix, indicating a need for either designing a problem tailored mixer layers in the QAOA ansatz or devising an improved encoding of the problem. 

Furthermore, the classical optimization subroutine of the QAOA responsible for estimating the optimal variational parameters, in general, is not uniquely determined, meaning that prior to implementing the problem on a quantum circuit, the choice of routine has to be optimized with respect to problem-specific characteristics. Now, as QAOA was originally intended as a general framework for solving optimization problems on near-term quantum devices, these artifacts pose an immediate challenge with regards to utilizing QAOA for combinatorial optimization.

The QAOA ansatz for the MSA cost-hamiltonian developed here was both simulated using Qiskit's software framework and run on Rigetti's Aspen M-3, both in combination with the classical COBYLA algorithm. While the simulation displays an increasing probability of sampling the feasible states, when increasing the layer repetition parameter $p$, with the state corresponding to the ideal gap arrangement of the MSA matrix being the most probable for $p\geq 4$, the results from Aspen M-3 exhibited a fairly significant degree of randomness, indicating the presence of an unsuitable amount of noise, and the need for utilizing a more specialized circuit-compiling strategy. 

Ultimately, this entails that the ability to efficiently utilize the QAOA ansatz on near-term devices to determine ideal MSAs with respect to the SP-scoring scheme is dependent on gaining further insight into the design of operators capable of restricting the mixing layer to the feasible subspace $\mathcal{S}$, and whether they can be efficiently decomposed and implemented in a gate based device for generally sized instances of MSA, alongside devising a more general strategy for the pre-processing optimization of the hyper-parameters.

\end{spacing}

\acknowledgements
The authors acknowledge fruitful discussions with Magnus Linnet Madsen and Marco Majland during various stages of this work. This work was supported in part by the Independent Research Fund Denmark, in part by the Novo Nordisk Foundation through grant number NNF20OC0065479, and in part by the European Innovation Council through Accelerator grant no. 190124924.


\bibliography{references}


\appendix
\section{Complexity analysis of MSA}
\label{sec: Combinatorial analysis of MSA}
\subsection*{Number of feasible solutions}

\begin{spacing}{\myspacing}
Consider a set of $N$ strings $\{\vec{s}_1,\vec{s}_2,...,\vec{s}_N\}$ of lengths $\{l_i\}$ with:
\begin{equation}
    L=\max\{l_i\}_{i=1}^N.
\end{equation}
Say that we want to fill each of the $N$ strings with gaps such that each of them conforms to length $L$, effectively restricting the number of gaps $g_i$ available for insertion in the $i'th$ string to:
\begin{equation}
    g_i = L-l_i.
\end{equation}
Because of the fact, that any valid MSA has to preserve the initial ordering of characters, it becomes apparent that the cardinality $\big|\mathcal{S}\big|$ of the space $\mathcal{S}$ of feasible solutions, i.e., the number of valid MSAs, for any instance of sequences, can be determined by simple counting, as the problem consequently amounts to counting the number of distinct ways to arrange $g_i$ identical gaps for $N$ different strings.

Particularly, for each string, the first of the $g_i$ identical gaps can be placed in $L$ different spots, the next in $L-1$, and so forth. Hence, the $g_i$ gaps can be arranged in
\begin{equation}
    L(L-1)(L-2)\hdots\big(L-(g_i-1)\big)=\frac{L!}{(L-g_i)!}
\end{equation}
ways. However, as the gaps are identical objects, symmetry reduces this number to the well-known binomial coefficient:
\begin{equation}
   L(L-1)(L-2)\hdots\big(L-(g_i-1)\big)\frac{1}{g_i!}=\frac{L!}{(L-g_i)!g_i!}=\binom{L}{g_i}
    \label{eq: nr alignemnts for 2}
\end{equation}
As $g_i=0$ for the string with $L_i=L$, we effectively 'only' have to worry about the arrangement of the $N-1$ strings. However, as:
\begin{equation}
    \binom{L}{L}=1,\quad \forall L \in\N.
\end{equation}
The number of feasible solutions can be written as:
\begin{equation}
    \big|\mathcal{S}\big|=\binom{L}{g_1}\binom{L}{g_2}\hdots\binom{L}{g_{N}}=\prod_{i=1}^{N}\binom{L}{g_i}
    \label{eq: cardinality of S}.
\end{equation}

\subsection*{Locally upper bounding fraction of feasible states}

Following the definition from the previous section, let us establish the fact that the fraction of feasible states:
\begin{equation}
    F\equiv\frac{|\mathcal{S}|}{|\mathcal{H}|}=\frac{\prod_{i=1}^{N}\binom{L}{l_i}}{2^{L\sum_{i=1}^Nl_i}},
\end{equation}
is at least exponentially decaying in terms of both the length of the reference string $L=\max_i(l_i)_i$, and the number of strings $N$. Initially, we define:
\begin{equation}
F_i = \frac{|\mathcal{S}_i|}{|\mathcal{H}_i|}=\frac{\binom{L}{l_i}}{2^{L l_i}},\,\,\,\textrm{such that}\,\,\, F=\prod_{i=1}^NF_i,
\end{equation}
and utilize the general fact that for $1\leq B\leq A$:
\begin{equation}
    \binom{A}{B} \leq \frac{A^B}{B!}
\end{equation}
such that:
\begin{equation}
    F_i \leq \frac{L^{l_i}}{l_i!}\frac{1}{2^{Ll_i}}=\frac{1}{l_i!}\Big(e^{\ln(L)}\Big)^{l_i}\Big(e^{\ln(2)}\Big)^{-Ll_i}=\frac{1}{l_i!}e^{-l_i[\ln(2)L-\ln(L)]}.
\end{equation}
Furthermore, in the smallest limiting case, all $l_i=1$ except the one reference string for which $l_i=L$, and as such:
\begin{equation}
    \sum_{i=1}^Nl_i=L+\sum_{i=1}^{N-1}l_i\geq L+N-1,
\end{equation}
and:
\begin{equation}
    \prod_{i=1}^{N}l_i !=L!\prod_{i=1}^{N-1}l_i!\geq L!.
\end{equation}
Therefore it is given that:
\begin{align}
    F&\leq \prod_{i=1}^N\frac{1}{l_i!}e^{-l_i[\ln(2)L-\ln(L)]}\\
     &\leq \frac{1}{L!}e^{-[\ln(2)L-\ln(L)][L+N-1]} \\
     &=e^{-\ln(L!)-[\ln(2)L-\ln(L)][L+N-1]}, \quad N,L \in \N_{\geq 2}.
\end{align}
The fraction of feasible states is upper bounded by a function that is super-exponentially decreasing in the length of the reference string $L$, and exponentially decreasing in the number of strings $N$.

\section{QUBO to Ising}
\label{sec: QUBO to ising}

Utilizing the $\{1,0\}\rightarrow\{-1,1\}$ variable change, yields:
\begin{align*}
    C(\vec{x})&=\sum_{i=1}^n\sum_{j=1}^nQ_{i,j}x_ix_j+\sum_{i=1}^nh_ix_i+d \nonumber \\
    &\quad\quad\Downarrow \\
    C(\vec{s})&=\sum_{i=1}^n\sum_{j=1}^nQ_{i,j}\bigg(\frac{1-s_i}{2}\bigg)\bigg(\frac{1-s_j}{2}\bigg)+\sum_{i=1}^nh_i\bigg(\frac{1-s_i}{2}\bigg)+d \nonumber\\
    &=\sum_{i=1}^n\sum_{j=1}^n\frac{Q_{i,j}}{4}+\sum_{i=1}^n\sum_{j=1}^n\frac{Q_{i,j}}{4}s_is_j-\sum_{i=1}^n\sum_{j=1}^n\frac{Q_{i,j}}{4}s_i-\sum_{i=1}^n\sum_{j=1}^n\frac{Q_{i,j}}{4}s_j+\sum_{i=1}^n\frac{h_i}{2}-\sum_{i=1}^n\frac{h_i}{2}s_i+d,
\end{align*}
which by symmetry yields the familiar grouping:
\begin{align*}
    C(\vec{s})&=\Bigg(\sum_{i=1}^n\sum_{j=1}^n\frac{Q_{i,j}}{4}s_is_j\Bigg)-\Bigg(\sum_{i=1}^n\sum_{j=1}^n\frac{Q_{i,j}+Q_{j,i}}{4}s_i+\sum_{i=1}^n\frac{h_i}{2}s_i\Bigg)+\Bigg(\sum_{i=1}^n\sum_{j=1}^n\frac{Q_{i,j}}{4}+\sum_{i=1}^n\frac{h_i}{2}+d\Bigg).
\end{align*}
Finally, the Ising formulation of (\ref{eq: general QUBO}) is obtained by vectorization, as:
\begin{equation*}
    C(\vec{s})\,\,\,=\underbrace{\frac{1}{4}\vec{s}^T\vec{Q}\vec{s}}_{\text{Quadratic term}}-\,\,\,\underbrace{\frac{1}{4}\bigg({\mathbbm{1}_n}^T\vec{Q}^T+{\mathbbm{1}_n}^T\vec{Q}+2\vec{h}^T\bigg)\vec{s}}_{\text{Linear term}}\,\,\,+\,\,\,\underbrace{\frac{1}{4}{\mathbbm{1}_n}^T\vec{Q}{\mathbbm{1}_n}+\frac{1}{2}{\mathbbm{1}_n}^T\vec{h}+d}_{\text{Constant term}},\,\,\,\,\,\,
\end{equation*}
where $\mathbbm{1}_n$ denotes an $n-$length vector of ones.

\section{Circuit}
\label{sec: circuit}

\begin{figure}[H]
    \centering
    \includegraphics[width=\linewidth]{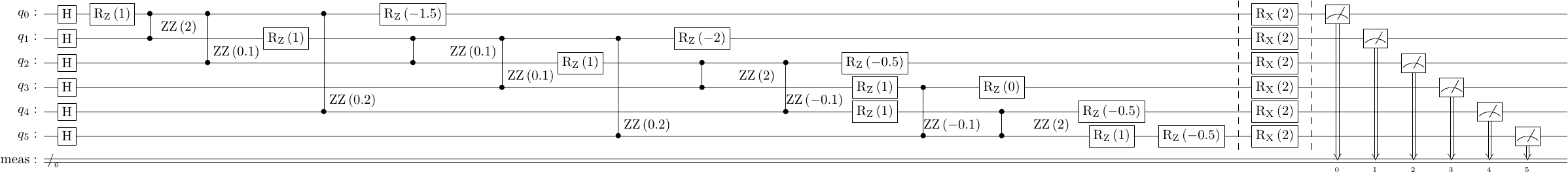}
    \caption{$p=1$ circuit for QAOA ansatz on MSA toy model from eq. (\ref{eq: Simple MSA toy model})}
    \label{fig:circuit}
\end{figure}

\end{spacing}

\end{document}